\documentclass[10pt,journal]{IEEEtran}
\usepackage{graphicx}          
\usepackage{amsmath}           
\usepackage{amssymb}
\usepackage{algorithm}
\usepackage{algpseudocode}
\usepackage{color}
\usepackage{subfigure}
\usepackage{cite}
\usepackage{enumerate}

\usepackage{enumitem}
\usepackage{epstopdf}
\usepackage[latin1]{inputenc}  

\usepackage{amsthm}

\theoremstyle{definition}

\theoremstyle{remark}
\theoremstyle{plain}

\theoremstyle{remark} 

\DeclareMathOperator{\E}{E}

\DeclareMathOperator*{\argmin}{arg\,min}
\DeclareMathOperator*{\argmax}{arg\,max}
\newcommand{\mbf}[1]{\mathbf{#1}}
\newcommand{\mbs}[1]{\boldsymbol{#1}}
\newcommand{\what}[1]{\widehat{#1}}

\newcommand{\y}{\mbf{y}}
\newcommand{\I}{\mbf{I}}
\newcommand{\0}{\mbf{0}}

\newcommand{\pvec}{\mbs{\theta}}

\newcommand{\ppos}{\mbf{x}}
\newcommand{\pclock}{\mbf{c}}
\newcommand{\ranges}{\mbs{\rho}}
\newcommand{\gradrange}{\mbs{\Gamma}}
\newcommand{\regclock}{\mbf{H}}
\newcommand{\regpos}{\mbf{G}}

\newcommand{\covnoise}{\mbf{Q}}
\newcommand{\IM}{\mbs{\Lambda}}
\newcommand{\FIM}{\mbf{J}}
\newcommand{\pder}{\partial}

\newcommand{\grange}{\mbs{\gamma}}

\newcommand{\T}{\top}
\newcommand{\proj}{\mbs{\Pi}^\perp_{\regclock}}
\newcommand{\W}{\mbf{W}}
\newcommand{\vvec}{\mbf{w}}
\newcommand{\pdir}{\mbf{p}}

\begin{document}

\title{Scalable and Passive  Wireless Network Clock Synchronization}
\author{Dave Zachariah, Satyam Dwivedi, Peter Händel and Petre Stoica\thanks{This
    work has been partly supported by the Swedish Research Council
    (VR) under contracts 621-2014-5874.}}

\maketitle

\begin{abstract}
Clock synchronization is ubiquitous in wireless systems for communication,
sensing and control. In this paper we design a scalable system
in which an indefinite number of passively receiving wireless units
can synchronize to a single master clock at the level of discrete
clock ticks. Accurate synchronization requires an estimate of the node positions. If such information is available the framework developed here takes position uncertainties into account. In the absence of such information we propose a mechanism which enables simultaneous synchronization and positioning. Furthermore we derive the Cramer-Rao bounds for the system
which show that it enables synchronization accuracy at sub-nanosecond
levels. Finally, we develop and evaluate an online estimation method
which is statistically efficient.
\end{abstract}

\section{Introduction}

Time synchronization plays a key role in wireless communication,
sensing and control. Indeed, many wireless applications require upkeep of
timing in accomplishing their objectives. 

In wireless cellular communications, accurate time information is traditionally needed for signal
acquisition, demodulation, multiple access coordination, etc
\cite{meyr1997digital, viterbi1995cdma}. Accurate timing and
synchronization are also requirements in real-time wireless channel
characterization and in several concepts in wireless communications, including
beamforming and interference alignment \cite{4785387, 6895306,
  6208110, 6472197}. Such requirements are also mentioned in
\cite{4785387} as the main challenge for distributed beamforming to work in the next generation wireless communication systems.
Similarly, in \cite{6895306, 6208110, 6472197}, accurate time synchronization is
shown to be a requirement for interference alignment to work. Emerging concepts like femto-cells pose more challenging synchronization
requirements in terms of scalability and accuracy as discussed in \cite{4623708}.
The sub-nanosecond time and phase synchronization
is also needed in distributed radar applications \cite{5259181}. 
Wireless ranging and positioning require time synchronization in time-difference-of-arrival (\textsc{Tdoa}) based schemes, where anchor nodes are synchronized in
time \cite{1343935}. Wireless control networks are also critically dependent on synchronized sensors and actuators \cite{PajicEtAl2011_wireless}.

Variants of
the Network Time Protocol (\textsc{Ntp}) \cite{103043} and the Precision
Time Protocol (\textsc{Ptp}) \cite{5223605} constitute the most popular methods
for time reference and synchronization in wired
networks\cite{5706321}. The emergence of a variety of wireless
networks during the past decade has led to the development of wireless
time-synchronization protocols and localization schemes. The Reference Broadcast
Synchronization (\textsc{Rbs}) and the Time
synchronization Protocol for Sensor Networks (\textsc{Tpsn}) emerged as popular
wireless time synchronization protocols around the same time
\cite{ElsonEtAl2002_fine, TPSN_ref}. In \textsc{Rbs}, the nodes in a wireless network synchronize through
a broadcast by a master node and inter-node exchanges to remove any
sender uncertainty. \textsc{Tpsn} works by creating a hierarchical tree-based
structure where every leaf node synchronizes to its parent node
through message exchanges. Neither \textsc{Rbs} nor \textsc{Tpsn} accounts for
propagation delays nor do they enable passive synchronization. For the aimed accuracies of these protocols, the
signal time-of-flight over a wireless channel is assumed to be
negligible. The protocol developed in \cite{Han2004_tsync} enables
higher accuracy by using separate channels for communication and
measurements required for synchronization.

\begin{figure}
  \begin{center}
    \includegraphics[width=0.70\columnwidth]{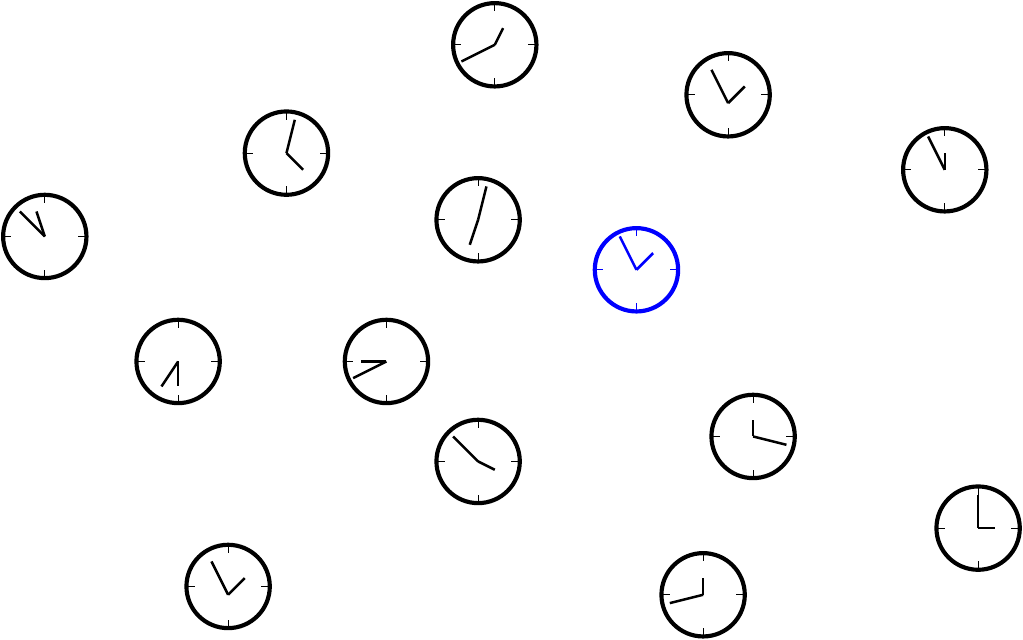}
  \end{center}
  \caption{Wireless network of nodes with local  clocks. The
    highlighted clock (blue) is a transmitting master unit to which 
    all passively receiving units should synchronize.}
  \label{fig:system_w_prior}
\end{figure}

Global Positioning System (\textsc{Gps}) signals are also used for synchronizing
time in wireless communication systems \cite{4623708, 5259181,
  6472197, anderson2009gps, 925191}. \textsc{Gps}-based timing solutions enable
an indefinite number of nodes to perform simultaneous self-localization
and time synchronization. This joint feature is important in 
deployed wireless sensor networks where both position and time need to
be resolved at each sensing node. In \textsc{Gps}-based solutions, a signal known as
pulse per second (\textsc{Ppp}) is extracted from pseudorange measurements and
satellite ephemeris data at the \textsc{Gps} receivers. The \textsc{Ppp} signal is then
used as a reference in frequency synthesizers to generate high
frequency signals \cite{6208110, 5259181, osterdock1995gps }. However,
\textsc{Gps} signals cannot be accessed indoors and the timing accuracy
obtained does not reach nanosecond levels.

In this paper we develop a scalable system in which passive,
receiver-only nodes
can synchronize to a single master clock at the level of discrete
clock ticks. We show that the synchronization performance of the system can
reach sub-nanosecond levels. When position information is lacking, we
propose a mechanism which enables simultaneous synchronization and
positioning at each node using three additional transceivers.

\subsection{Prior art and our contributions}

Time synchronization
schemes are evolving to provide nanosecond-level 
synchronization, which requires accounting for signal time-of-flight
between nodes. A scalable multihop scheme to
synchronize the nodes to nanosecond accuracy was proposed in
\cite{SeguraEtAl2014_experimental}.

Several works have developed system proposals as well as presented
theoretical analyses of time synchronization, cf. \cite{FrerisEtAl2010_fundamentals,
  EtzlingerEtAl2014_cooperative, NohEtAl2008_new}. Fundamental
limits on time synchronization in sensor networks were given in
\cite{FrerisEtAl2010_fundamentals}. The authors of
\cite{EtzlingerEtAl2014_cooperative} suggested using factor-graph 
methods for network clock estimation. In \cite{NohEtAl2008_new},
clock synchronization is achieved using eavesdropping
measurements. The synchronizing unit is a receiver-only node and hence
the method is claimed to be energy efficient. In
\cite{Weiss2014_passive} a joint localization method for 
source nodes was proposed using \textsc{Tdoa} which implicitly synchronizes an arbitrary
number of anchor nodes. 




Our proposed method for clock synchronization in wireless networks enables system
performance beyond the state-of-the-art. Specifically, we highlight the
following attributes of our proposal.
\begin{itemize}
\item \textit{Accuracy:} We focus on enabling nanosecond accuracy. \textsc{Ntp} provides millisecond accuracy over IP networks and has
  been overtaken by \textsc{Ptp} over wired networks. \textsc{Ptp} provides accuracy
  levels of a few nanoseconds using specialized hardware. For wireless solutions, 
such as \textsc{Rbs} and \textsc{Tpsn}, the accuracy for sensor network synchronization
methods is on the order of microseconds.  These methods do not need to take
time of flight into account as their requirements are less
stringent. \textsc{Gps}-based synchronization methods can typically synchronize
to a  100 nanosecond-level. For nanosecond levels, time of flight needs to be
estimated accurately as in
\cite{SeguraEtAl2014_experimental}. 

\item \textit{Scalability:} Another feature of the proposed synchronization method is its
  scalability. Scalability has been addressed previously in a few papers,
  albeit only implicitly. \textsc{Rbs}, \textsc{Tpsn} and the scheme proposed in
  \cite{SeguraEtAl2014_experimental} are scalable by virtue of
  providing synchronization to nodes through adhoc multihop
  connections. In these systems, nodes synchronize through mutual
  exchanges of signals among them. The signal exchange could be two-way
   round-trip time measurements or timestamps
   recording time-of-arrival information. By contrast, systems like \textsc{Gps} and the one proposed in
  \cite{NohEtAl2008_new} are receiver-only systems and hence they
  allow any number of nodes to synchronize with the reference
  clock. Our proposed method is similar to the latter class of scalable
  solutions. Indeed, we develop a method that is
  scalable as each synchronizing node requires only a receiver to
  sychronize to a reference, as in \textsc{Gps}. The lack of transmission
  requirement for the synchronizing nodes makes the solution energy
  efficient. 

\item \textit{Positioning for synchronization:} Our proposed solution
  is similar to \textsc{Gps} with respect to scalability but enables nanosecond
  accuracy using existing hardware technologies. In addition, it can
  be used in indoor scenarios. 
We will propose a local positioning system along with the synchronization
mechanism to enable time-of-flight estimation. The proposed
positioning system for synchronization builds upon our previous works
\cite{6340491, 6510560, 6490337, Zachariah2014}. 
\end{itemize}

We consider a general scenario as illustrated in Figure~\ref{fig:system_w_prior}.
The observed clock time in a wireless network is traditionally
modeled as a continuous function of clock skew $\alpha$ and the phase offset $\beta$ \cite{NohEtAl2008_new,Skog&Handel2010_synchronization,WuEtAl2011_clock},
\begin{equation}
C_m(t) = t \ \  \mbox{and} \ \ C_u(t) = \alpha t + \beta,
\end{equation}
where $C_m(t)$ denotes a reference master clock and $C_u(t)$ denotes
the local clock of a node $u$ in the network. In this model the local
clock time can be resolved into that of the master clock by
identifying the clock parameters. Network synchronization is achieved by resolving the observed time at
each node to a common clock.

In digital clocks, however, time is recorded by counting the number of periods of a
repeating clock signal. At each rising clock edge of the periodic
signal, an integer time counter is incremented. Our goal is to resolve the time
observed on clocks at nodes $u=1, \dots, U$. To achieve resolution levels
below that of the clock period we propose using a time measuring device that can
observe \emph{intervals} between discrete time events. Such events are
defined as periodic ticks on the digital clock and as received
signals from the master node $m$.

More concretely, to enable sub-nanosecond accuracy in time synchronization, we propose
the usage of:
\begin{itemize}
\item High bandwidth signals. As is widely documented in the literature, the precision of time of arrival
  measurements is inversely proportional to the square  bandwidth of the transmitted
  signal \cite{4802196, falsi2006time}.
\item Accurate time-interval measurement device. Examples include high speed
  analog-to-digital (\textsc{Adc}) converters and time-to-digital converters
  (\textsc{Tdc}); such devices can measure time intervals with
  sub-nanosecond accuracy. In \cite{SeguraEtAl2014_experimental} a high speed \textsc{Adc} was
  used with sampling frequency greater than $1$\,  Giga samples per second. In \cite{DwivediEtAl2015_jointsyncrange}, clock parameter
  estimation for two clocks was experimentally demonstrated using a
  \textsc{Tdc} with a precision of nearly $100$\, picoseconds.
\end{itemize}

The solution proposed in \cite{Weiss2014_passive} is a recent, novel way of synchronizing fixed anchor nodes while estimating the 
positions of several emitting source nodes. In our setup, passive
nodes with unknown positions can synchronize to a master node.












\begin{figure}[!h]
  \begin{center}
    \includegraphics[width=0.8\columnwidth]{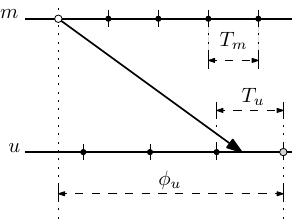}
  \end{center}
  \caption{Space-time diagram of nodes $m$ and $u$, with one vertical spatial dimension and a horizontal time dimension. The digital clock
    states correspond to discrete events or ticks along the time-axes
    (dots). The master node $m$ and passive node $u$ have clock periods
    $T_m$ and $T_u$, respectively. The transmission event from the
    master defines the initial tick of the system  clock (white).
 Upon receiving the signal, the corresponding initial tick on the local clock
 (gray) will be subject to an unknown offset $\phi_u$.}
  \label{fig:masterclock_basic}
\end{figure}

\subsection{Problem formulation}

The state of each digital clock is the integer number of 
cycles that have elapsed since some initialization event.
Suppose the master clock operates with a period $T_m$. Then its
clock state $n_m \in \{ 0, 1, 2, \dots \}$ corresponds to times
$$\mathcal{C}_m \in \{ 0, \; T_m, \; 2T_m, \; \dots \}.$$
The master clock initializes the counters by transmitting a signal
across the wireless network. The clock at node $u$, which operates with period
$T_u$, will have a relative offset $\phi_u$ due to the propagation delay and
nonsynchronicity as illustrated in
Fig.~\ref{fig:masterclock_basic}. Its
clock state $n_u \in \{ 0, 1, 2, \dots \}$ corresponds to times
$$\mathcal{C}_u \in \{ \phi_u, \; T_u + \phi_u, \; 2T_u + \phi_u, \; \dots \}.$$
Therefore the current clock state $n_u$ of the node can be resolved
into a common time if the clock parameters $\phi_u$ and $T_u$ are
identified. In addition, identification of $T_m$ enables also
coordination with respect to the master periodic signal across the wireless network.

Based on the previous discussion we may write
\begin{equation}\label{eq:clocknetwork}
\mathcal{C}_m = T_m n_m \quad \text{and} \quad
\begin{cases}
\mathcal{C}_1 &= T_1 n_1 + \phi_1 \\
\mathcal{C}_2 &= T_2 n_2 + \phi_2 \\ 
&\vdots \\
\mathcal{C}_U&= T_U n_U + \phi_U \\
\end{cases}.
\end{equation}
By identifying the clock
parameters at each node $u=1,\dots,U$, synchronization is achieved since a
common time frame is shared across the entire network. This enables
coordination relative to the master clock among all nodes.

Note that nominal values of the clock frequencies, and therefore of the
periods $T_u$ and $T_m$, are typically available given. However, usually, these values are not sufficiently precise. To obtain more accurate
estimates of $T_u$ it is possible to use a device that measures the intervals between ticks. Similarly, as the signal from the
master clock is repeated periodically after $M$ cycles, $T_m$ can also be
estimated accurately. The primary challenge, however, is to estimate
the relative offset $\phi_u$.

In this paper, we design a system in which passively receiving nodes are
synchronized by estimating their respective clock parameters. The system
is scalable to an indefinite number of nodes, i.e. $U \gg 1$. Furthermore, we study the resolution limits of the system
using the Cramér-Rao bounds. Using existing hardware performance
figures, we show that the proposed system enables sub-nanosecond
accuracy. While the estimation of $T_m$ and $T_u$ can be performed
separately from $\phi_u$, we derive a joint online estimator that
takes into account the uncertainties of all estimates. The proposed estimator
is subsequently evaluated in several numerical experiments.

\emph{Remark:} An implementation of the estimator along with numerical
simulation examples is available at the webpage of KTH Dept. Signal
Processing under `Reproducible research'.


\section{System model}
\label{sec:systemmodel}
 
\begin{figure}
  \begin{center}
    \includegraphics[width=0.70\columnwidth]{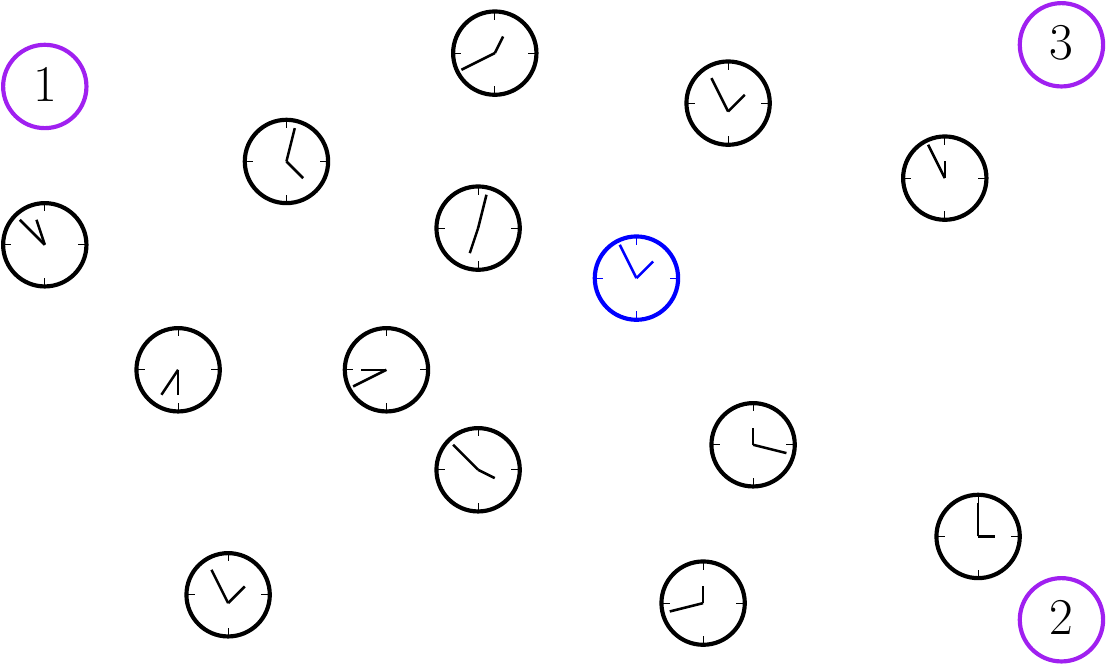}
  \end{center}
  \caption{System model with three additional transceivers at known positions.}
  \label{fig:system_w_transceivers}
\end{figure}

To achieve the objectives stated above, we propose a system with
the following features:
\begin{enumerate}
\item All passive units can measure time-intervals $\Delta = t - t'$
  between events at times $t$ and $t'$, using a time measurement
  device. This enables observations at a higher resolution than that of the
  digital clock and is grounded in the emerging \textsc{Tdc} and \textsc{Adc} technologies.

\item The master periodically transmits a
 time-resolvable signal after $M$ clock cycles. Among others, this ensures the
  identifiability of $T_m$. The transmission event from the master defines the starting
  point of a system-wide clock with period $T_m$. We call the period of $M$ clock cycles an \emph{epoch}. 

\item  The master node $m$ is located at a known position
  $\ppos_m$. The position of an arbitrary synchronizing node
  $u$, denoted $\ppos$, is unknown. Together with the assumption that an epoch is longer than the
clock period of any synchronizing node, i.e., $MT_m > T_u$, that fact that $\ppos_m$
is known enables the identifiability of $\phi_u$ as we will show below.
\end{enumerate}
We will model the unknown position as $\ppos \sim
\mathcal{N}(\bar{\ppos}, \IM^{-1}_x)$ when we have access to a prior
estimate $\bar{\ppos}$ with a dispersion matrix 
$\IM^{-1}_x$. When such prior position information is lacking,
i.e. when $\IM_x = \0$, then $\phi_u$ cannot be identified. To
ensure identifiability in such a case, under the assumption that the positions are expressed in three-dimensional coordinates, we consider a system with the following additional features:
\begin{enumerate}[resume]
\item There exists three transceiving nodes, deployed at known positions $\{
  \ppos_1,  \ppos_2,  \ppos_3 \}$, cf. Fig.~\ref{fig:system_w_transceivers}. The transceivers transmit sequentially in the order $\{m,1,2,3 \}$, and repeatedly.

\item When receiving a signal from the
  preceding transmitter in the above order, the subsequent transceiver transmits after a
  fixed delay $\Delta_0$, which can be generated
  independently of the local clock \cite{Zachariah2014}. This is to avoid
  interfering signals from the master and transceivers during an
  epoch. Specifically, we assume
\begin{equation*}
MT_m \gg \Delta_0 > \text{max. distance to transmitter} / c,
\end{equation*}
where $c$ is the propagation velocity. Then each
transmitted signal can reach all nodes before the subsequent signal is transmitted. 
\end{enumerate}

Making use of these features together, we will show that it is possible to
synchronize any number of passively receiving nodes. That is, each
synchronizing node $u$ can resolve the unknown clock parameters $\phi_u$,
$T_u$ and $T_m$ in \eqref{eq:clocknetwork}. The \emph{s}calable \emph{wi}reless \emph{n}etwork \emph{s}ynchronization system is abbreviated \textsc{Swins}.

\subsection{Data model}

\begin{figure}
  \begin{center}
    \includegraphics[width=0.8\columnwidth]{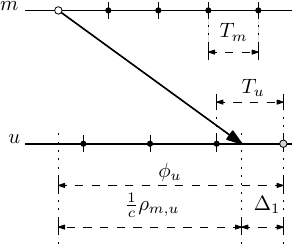}
  \end{center}
  \caption{Space-time diagram of nodes $m$ and $u$. $\Delta_1$ is defined as the time interval
between the received signal and the subsequent clock tick at
$u$ (gray). The time of flight equals $\frac{1}{c} \rho_{m,u}$.}
  \label{fig:masterclock}
\end{figure}

First, consider
the initial signal received by a passive node $u$ from $m$, as depicted
in Fig.~\ref{fig:masterclock}.
Node $u$ can only record time intervals, and we define $\Delta_1$ as the time
between the received signal and the next clock tick at
$u$. Given that the time of flight of the signal is $\frac{1}{c}
\rho_{m,u}$, where $c$ is the signal propagation velocity and
$\rho_{m,u} = \| \ppos_m - \ppos \|_2$ is the range between $m$
and $u$, the following relation 
\begin{equation}\label{eq:delta}
\Delta_1 = \phi_u - \frac{1}{c} \rho_{m,u}
\end{equation}
applies to the first epoch.

\begin{figure}
  \begin{center}
    \includegraphics[width=1.00\columnwidth]{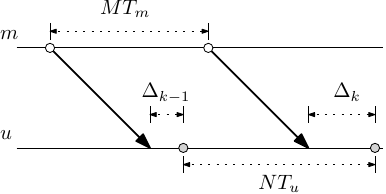}
  \end{center}
  \caption{Space-time diagram of nodes $m$ and $u$ over an epoch of
    $M$ clock cycles for the master clock (white). At the local clock, each $N$th clock cycle is observed (gray).}
  \label{fig:epoch}
\end{figure}

At node $u$, the number of clock cycles till the subsequent epoch begins, denoted $N$,
is recorded and corresponds to a constant time interval $NT_u \geq
MT_m$. Observing each $N$th clock tick we can derive a relation between the observed intervals as follows, see Fig.~\ref{fig:epoch} which illustrates the basic principle. Let $\Delta_k$ denote the time between receiving a signal and the $kN$th clock tick for $k>1$. Then it follows that $\Delta_{k-1} + NT_u = MT_m + \Delta_k$.
This relation for $k$ intervals together with \eqref{eq:delta} can be written as the following recursion
\begin{equation*}
\begin{split}
\Delta_k &= \Delta_{k-1} + N T_u - M T_m \\
& \quad \vdots \\
\Delta_2 &= \Delta_1 + N T_u - M T_m \\
\Delta_1 &= \phi_u - \frac{1}{c} \rho_{m,u} 
\end{split}
\end{equation*}
which comprises the unknown clock and position parameters. Using this recursion, we can write the observed interval $\Delta_k$ at the $k$th epoch as 
\begin{equation}\label{eq:measurement_phi}
\begin{split}
y_{\phi,k} &=\phi_u - \frac{1}{c} \rho_{m,u} + (k-1)(N T_u - MT_m) + w_{\phi,k},
\end{split}
\end{equation}
where $w_{\phi,k}$ is a zero-mean noise. From the above equation we
see that $\phi_u$ cannot be identified without determining also the range
$\rho_{m,u}$ which is a function of the unknown position
$\ppos$. 

\begin{figure}
  \begin{center}
    \includegraphics[width=1.0\columnwidth]{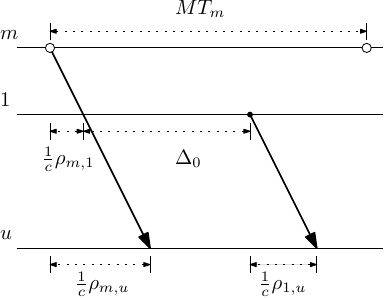}
  \end{center}
  \caption{Space-time diagram of nodes $m$, $1$ and $u$ over an
    epoch. Upon receiving a signal, the transceiving node $1$ transmits its signal
    after a known delay $\Delta_0$. The interval between the received
    signals at $u$ is: $\frac{1}{c}\rho_{m,1} + \Delta_0 + \frac{1}{c}\rho_{1,u} -
\frac{1}{c}\rho_{m,u} $. Note that the apparent congruence with $\Delta_0$ is a coincidence of the example in one-dimensional space and does not hold in general.}
  \label{fig:fullcycle}
\end{figure}

Next, we show that it is possible to resolve $\ppos$ using scheduled
transmissions from the three transceivers \emph{during} an epoch. The
basic principle is illustrated in Fig.~\ref{fig:fullcycle}. When the
master signal reaches transceiver node 1, it transmits after a known delay $\Delta_0$. 
The subsequent transceiving nodes do the same according to the given transmission order $\{m,1, 2, 3 \}$.
For the $k$th epoch, the time-intervals between each received signals at node $u$ can be written as 
\begin{equation}\label{eq:measurement_rho}
\begin{split}
y_{1,k} &= \frac{1}{c}\rho_{m,1} + \Delta_0 + \frac{1}{c}\rho_{1,u} -
\frac{1}{c}\rho_{m,u} +  w_{1,k}, \\
y_{2,k} &= \frac{1}{c}\rho_{1,2} + \Delta_0 + \frac{1}{c}\rho_{2,u} -
\frac{1}{c}\rho_{1,u} +  w_{2,k}, \\
y_{3,k} &= \frac{1}{c}\rho_{2,3} + \Delta_0 + \frac{1}{c}\rho_{3,u} -
\frac{1}{c}\rho_{2,u} +  w_{3,k} ,
\end{split}
\end{equation}
where $\rho_{i,j} = \| \mbf{x}_i - \mbf{x}_j \|_2$. Each time-interval
measurement produces a hyperbolic constraint on
$\mbf{x}$, cf. the principles of \textsc{Tdoa} approach \cite{Zachariah2014}. Thus three constraints are sufficient for identifying
$\mbf{x}$ in the three-dimensional space, and therefore also for resolving $\phi_u$.

At the end of the $k$th epoch, its duration is recorded, resulting in 
\begin{equation}\label{eq:measurement_Tm}
\begin{split}
y_{m,k} &= M T_m  + w_{m,k},
\end{split}
\end{equation}
where $M$ is known and therefore we can resolve $T_m$ from \eqref{fig:epoch}. Similarly, for each epoch at $u$,
$N$ ticks are recorded at the local clock, and the observed time interval is
\begin{equation}\label{eq:measurement_Tu}
\begin{split}
y_{u,k} &= N T_u + w_{u,k}. \\
\end{split}
\end{equation}

In sum, using the above observations, made at a passive node $u$, ensures that
$\phi_u$, $T_u$ and $T_m$ are identifiable parameters. This enables
wireless synchronization to the master clock $m$. The additional transceivers also render $\ppos$ identifiable and therefore enable self-localization at each node $u$. 

\subsection{Noise model}\label{sec:noisemodel}

Each observed time-interval above is subject to two sources of error
arising from its start and stop events, respectively. In
\eqref{eq:measurement_rho} and \eqref{eq:measurement_Tm}, the start and
stop events are triggered by uncorrelated RF signals. A nominal value of error
variance $\sigma^2_0$ from such events can be assigned but in practice
varying RF conditions produce outliers so that we assume a
varying $\sigma^2_k$. The total noise variance for these measured
intervals is $\E[w^2_{i,k}] = 2\sigma^2_k $ for $i=m, 1, 2, 3$ since
they are based on a pair of RF measurements.
 In \eqref{eq:measurement_Tm}, one RF
measurement is shared with \eqref{eq:measurement_phi} so that the errors of the observed intervals are correlated: $\E[
w_{\phi,k} w_{m,k} ] = \sigma^2_k$. Furthermore, because two consecutive
measurements share one RF measurement in \eqref{eq:measurement_Tm} and
\eqref{eq:measurement_rho}, we can write:
\begin{equation*}
\E[w_{i,k} w_{j,k}]=\begin{cases} 2\sigma^2_k, & i=j \\
\sigma^2_k, & j \text{ follows } i, \text{ or vice versa,}  \\
0, & \text{ otherwise.}
\end{cases}
\end{equation*}

We assume that the RF noise yields the dominant part of $\sigma_k$ and that the noise
contribution of the timing device itself is only a small fraction $0 <
\alpha <1 $ of $\sigma_k$, which depends
on the performance figures of the device. In practice $\alpha = 0.1$
is a reasonable value for existing hardware
\cite{DwivediEtAl2015_jointsyncrange}, and this is the value we will
assume in what follows. Then since the start and stop events of the interval in
\eqref{eq:measurement_Tu} are triggered solely by two clock ticks, we
have $\E[w^2_{u,k}] = 2\alpha^2 \sigma^2_k$. Finally, beacause \eqref{eq:measurement_phi} is based on
one RF and one clock tick we have $\E[w^2_{\phi,k}] =(1+\alpha^2)\sigma^2_k$.
We model the noise sources as jointly Gaussian and omit the
correlation between consecutive epochs.

\section{Cramér-Rao bounds}

To study some basic properties of \textsc{Swins}, we begin by
collecting the observed time intervals from epoch $k$ in a vector
\begin{equation}\label{eq:y_k}
\y_k \triangleq \mbf{S}_k \begin{bmatrix} y_{\phi,k} & y_{u,k} &
  y_{m,k} & y_{1,k} & y_{2,k} & y_{3,k} \end{bmatrix}^\top \in \mathbb{R}^{n_k},
\end{equation}
where 
\begin{equation}
\mbf{S}_k = \begin{cases}
\I_6 & \text{if transceiving nodes present in epoch } k, \\
[\I_3 \; \0_{3\times 3}] & \text{otherwise} 
\end{cases}
\end{equation}
is a selection matrix and $n_k$ is the number of measured
intervals in epoch $k$. Combining \eqref{eq:measurement_phi},
\eqref{eq:measurement_Tu}, \eqref{eq:measurement_Tm}, and
\eqref{eq:measurement_rho}, we can write \eqref{eq:y_k} as
\begin{equation}\label{eq:vectormodel}
\mbf{y}_k =  \mbs{\mu}_k + \regclock_k \pclock +
\frac{1}{c}\regpos_k\ranges(\ppos) + \mbf{w}_k \in
\mathbb{R}^{n_k},
\end{equation}
where $\pclock \triangleq [  \phi_u  \; T_u \; T_m   ]^\top$ contains the
parameters of interest. The mean
vector
\begin{equation*}
\mbs{\mu}_k = \mbf{S}_k\begin{bmatrix}
0 \\ 0  \\ 0 \\ \frac{1}{c}\| \ppos_m -\ppos_1  \| + \Delta_0\\
\frac{1}{c}\| \ppos_1 -\ppos_2  \| + \Delta_0 \\ \frac{1}{c}\|
\ppos_2 -\ppos_3  \| + \Delta_0 \end{bmatrix} \in \mathbb{R}^{n_k}
\end{equation*}
is known and the vector of ranges 
\begin{equation*}
\ranges(\ppos)  = \begin{bmatrix}\| \ppos-\mbf{x}_m \|_2 \\
\|\ppos-\mbf{x}_1 \|_2 \\ 
\|\ppos-\mbf{x}_2 \|_2 \\
\|\ppos-\mbf{x}_3 \|_2\end{bmatrix} \in \mathbb{R}^4,
\end{equation*}
is a function of the unknown position $\mbf{x}$. The known system
matrices in \eqref{eq:vectormodel} can be written as
\begin{equation*}
\begin{split}
\regclock_k &=  \mbf{S}_k\begin{bmatrix} 1 & (k-1)N & -(k-1)M  \\0 & N & 0 \\0 & 0 & M \\ 0 & 0 & 0 \\  0 & 0 &0 \\ 0 & 0
  &0 \end{bmatrix},\\
\regpos_k &=  \mbf{S}_k\begin{bmatrix} -1 & 0 &
  0 & 0 \\ 0 & 0 & 0 & 0 \\ 0 & 0  & 0 & 0 \\ -1 & 1  & 0 & 0 \\  0 & -1
  & 1 & 0 \\ 0 & 0  & -1 & 1 \end{bmatrix}.
\end{split}
\end{equation*}

Based on the noise model introduced in Section~\ref{sec:noisemodel},
the measurement noise vector $\mbf{w}_k$ has a covariance matrix
$ \sigma^2_k \mbf{Q}_k \triangleq \E[\mbf{w}_k \mbf{w}^\top_k] $,
given by:
\begin{equation*}
\covnoise_k =  \mbf{S}_k\begin{bmatrix}
(1+\alpha^2) & 0 & 1 & 0 & 0 &0 \\
0 & 2\alpha^2 & 0 & 0 & 0 &0 \\
1 & 0 & 2 & 1 & 0 & 0 \\
0 & 0 & 1 & 2 & 1 & 0 \\
0 & 0 & 0 & 1 & 2 & 1\\
0 & 0 & 0 & 0 & 1 & 2
\end{bmatrix}\mbf{S}^\top_k.
\end{equation*}
 As the noise $\mbf{w}_k$ is modeled as Gaussian, we have 
\begin{equation}\label{eq:ydistribution}
\y_k | \pclock, \ppos, \sigma^2_k \sim \mathcal{N}( \mbs{\mu}_k + \regclock_k \pclock +
\frac{1}{c}\regpos_k\ranges(\ppos), \sigma^2_k \covnoise_k).
\end{equation}
This data model enables an analysis of how accurately the clock parameters can be estimated in \textsc{Swins}.

\subsection{Cramér-Rao bound}

Define the vector $$\pvec \triangleq \begin{bmatrix} \pclock \\
  \ppos\end{bmatrix}\in \mathbb{R}^{3+d},$$
where $d=2$ or $3$ is the spatial dimension. The Fisher information
matrix of $\pvec$ for the $k$th epoch data model in
\eqref{eq:ydistribution} is given by \cite[ch.~3]{Kay1993_ssp}\cite[App.~B.3]{Stoica&Moses2005_spectral}:
\begin{equation}\label{eq:FIM}
\FIM_k(\ppos,\sigma^2_k)  = \frac{1}{\sigma^2_k}
\left[\regclock_k \; \; \frac{1}{c}\regpos_k\gradrange(\ppos)\right]^\top \covnoise^{-1}_k \left[\regclock_k \; \; \frac{1}{c}\regpos_k\gradrange(\ppos)\right]
,
\end{equation}
where the Jacobian of the range function $\ranges(\ppos)$ is
\begin{equation*}
\gradrange(\ppos) \triangleq \partial_{x} \ranges(\ppos) = \begin{bmatrix}
  \frac{(\ppos-\ppos_m)^\top}{\| \ppos-\ppos_m \|_2} \\
  \frac{(\ppos-\ppos_1)^\top}{\| \ppos-\ppos_1 \|_2} \\
  \frac{(\ppos-\ppos_2)^\top}{\| \ppos-\ppos_2 \|_2} \\
  \frac{(\ppos-\ppos_3)^\top}{\| \ppos-\ppos_3 \|_2} \end{bmatrix} \in \mathbb{R}^{4 \times d}.
\end{equation*}

In the above model, the data from each epoch are mutually
uncorrelated. Therefore the information from each epoch is additive and the total information matrix after $k$ epochs equals
\begin{equation}\label{eq:sumFIM}
\begin{split}
\IM_{k} &= \IM_{k-1} + \FIM_k,
\end{split}
\end{equation}
where $\IM_{0} = \mbf{0}$. Then the mean-square error (\textsc{Mse}) matrix of
any unbiased estimator $\hat{\pvec}$ is bounded via the Cramér-Rao
inequality:
\begin{equation*}
\E_y[(\pvec - \hat{\pvec})(\pvec- \hat{\pvec})^\top] \succeq \IM^{-1}_k,
\end{equation*}
and specifically for  $\hat{\pclock}$ we have
\begin{equation}\label{eq:crb}
\E_y[(\pclock - \hat{\pclock})(\pclock - \hat{\pclock})^\top] \succeq (
 \IM_{c,k} - \IM^\top_{xc,k}  \IM^{-1}_{x,k} \IM_{xc,k} )^{-1},
\end{equation}
where the right-hand side is obtained by partitioning the information
matrix as 
\begin{equation*}
\IM_k =
\begin{bmatrix}
\IM_{c,k} & \IM^\top_{xc,k} \\
\IM_{xc,k}  & \IM_{x,k}
\end{bmatrix}.
\end{equation*}
Note that the information matrix, via \eqref{eq:FIM}, is 
dependent on $\ppos$ and $\sigma^2_k$, but not on the clock parameters $\pclock$.

\begin{figure}
  \begin{center}
    \includegraphics[width=1.0\columnwidth]{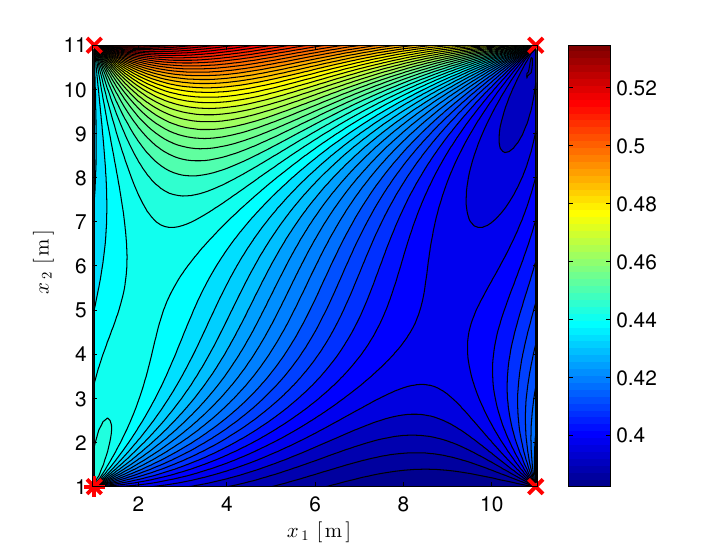}
  \end{center}
  \caption{Resolution limit of $\phi_u$ in [ns], using the square-root
    of the \textsc{Crb}, as a function of $\ppos$. The noise level $\sigma_k$ is fixed
    to 5~[ns] and 250 epochs are observed. The master and transceiver locations
    are denoted by an asterisk and by crosses, respectively. }
  \label{fig:crb_phi}
\end{figure}

To illustrate the spatial
dependence of $\IM_k$ on $\ppos$, for $d=2$, we plot the Cramér-Rao bound (\textsc{Crb}) of $\phi_u$ as a function of $\ppos$ in
Fig.~\ref{fig:crb_phi}. We set the known positions of the master and
transceivers as
\begin{equation}
\ppos_m = \begin{bmatrix} 1 \\ 1 \end{bmatrix}, 
\ppos_1 = \begin{bmatrix} 11 \\ 11 \end{bmatrix},
\ppos_2 = \begin{bmatrix} 1 \\ 11 \end{bmatrix},
\ppos_3 = \begin{bmatrix} 11 \\ 1 \end{bmatrix}.
\end{equation}
The noise standard deviation $\sigma_k$ is fixed to 5~[ns] and the fraction arising
from the timing device is set to $\alpha = 0.1$, which are reasonable
figures for existing hardware technologies. Using 250 epochs we see
that the resolution limit of \textsc{Swins} is on the order of sub-nanoseconds across space. We note in the passing that the spatial configuration of the transmitting nodes $\{m, 1, 2, 3\}$ and their transmission order result in a slightly lower limit in the bottom right quadrant.

\subsection{Hybrid Cramér-Rao bound}

When an informative prior for $\ppos$ exists, the unknown
position can be modeled as a random variable $\ppos \sim \mathcal{N}(\bar{\ppos},
\IM^{-1}_x)$. Then the MSE matrix of any unbiased estimator
$\hat{\pclock}$, when averaged over all possible values of $\ppos$, is
bounded via the Hybrid Cramér-Rao inequality \cite{vanTrees2004_detection}:
\begin{equation}
\E_{y,x}[(\pclock - \hat{\pclock})(\pclock - \hat{\pclock})^\top] \succeq (
\bar{\IM}_{c,k} - \bar{\IM}^\top_{xc,k}  \bar{\IM}^{-1}_{x,k} \bar{\IM}_{xc,k} )^{-1},
\label{eq:hcrb}
\end{equation}
where the right-hand side is obtained from the expected information
matrix
\begin{equation*}
\bar{\IM}_k =
\E_x[ \IM_k ] +
\begin{bmatrix} \0 & \0 \\ \0 & \IM_x \end{bmatrix} = 
\begin{bmatrix}
\bar{\IM}_{c,k} & \bar{\IM}^\top_{xc,k} \\
\bar{\IM}_{xc,k}  & \bar{\IM}_{x,k}
\end{bmatrix}.
\end{equation*}
The expectation is approximated numerically using Monte Carlo simulations.

\begin{figure}
  \begin{center}
    \includegraphics[width=1.0\columnwidth]{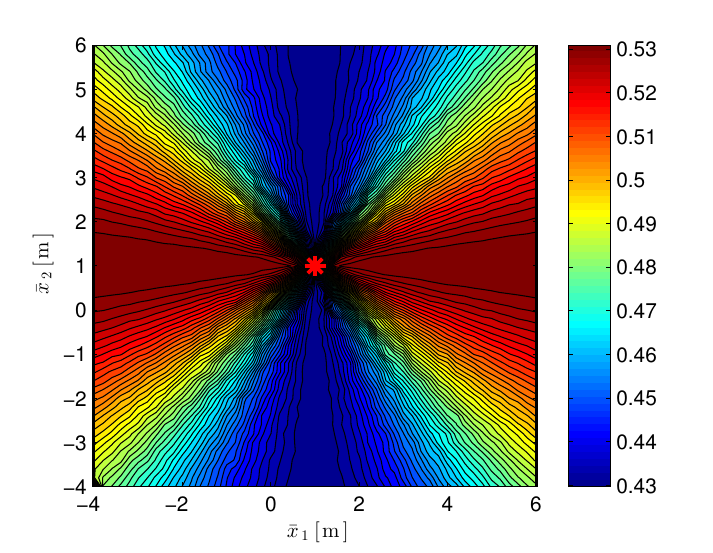}
  \end{center}
  \caption{Resolution limit of $\phi_u$ in [ns], using the square-root
    of the \textsc{Hcrb}, as a function of the
    prior mean $\bar{\ppos}$. The noise level $\sigma_k$ is fixed
    to 5~[ns] and 250 epochs are observed. The master location
    $\mbf{x}_m$ is denoted by an asterisk.}
  \label{fig:hcrb_phi}
\end{figure}

To illustrate the spatial
variation of \eqref{eq:hcrb} for $d=2$, we drop the transceiving nodes and plot in
Fig.~\ref{fig:hcrb_phi} the Hybrid Cramér-Rao Bound (\textsc{Hcrb}) of $\phi_u$
as a function of the prior mean $\bar{\ppos}$. The master position and
the precision matrix of the prior of $\ppos$ are given by:
\begin{equation*}
\ppos_m = \begin{bmatrix} 1 \\ 1 \end{bmatrix}, \quad
\IM_x = 
\begin{bmatrix}
0.1^2 & 0 \\
0 & 0.01^2
\end{bmatrix}^{-1}.
\end{equation*}
This corresponds to an position error ellipse whose axis correspond to standard deviations of 0.1 and 0.01 meters, respectively. There is greater uncertainty along the $\bar{x}_2$-axis than the $\bar{x}_1$-axis. Consequently the bound on the clock error, which depends on the range to the master, is greater when $\bar{\ppos}$ is on positions along one axis than the other. This variation in the resolution limit is clearly visible in  Fig.~\ref{fig:hcrb_phi}. Observe that the prior precision of $\ppos$ is sufficient to enable sub-nanosecond accuracy.

\section{Online estimator}

In this section, we derive an online estimator for the clock parameters $\pclock$ and position $\ppos$. The method refines the estimate at each epoch $k$. Its overall memory requirement is constant and computational complexity scales linearly with $k$.

\subsection{Linear combiner}

Our interest here is to process the data from each epoch $\y_k$
sequentially and then form a linear combination
of the so-obtained estimates. This combined estimate is recursively
computed and, as we will see, attains the Cramér-Rao bounds of the
system asymptotically.

The prior knowledge about the position can be
equivalently expressed as $\bar{\ppos} \sim
\mathcal{N}(\ppos, \IM^{-1}_x)$.
Then for epoch $k$, we can formulate the maximum likelihood estimate
\begin{equation}\label{eq:ML}
\check{\pvec}_k = \argmax_{\pvec} \; \left[ \max_{\sigma^2_k} \;
  p(\y_k, \bar{\ppos} | \pvec, \sigma^2_k) \right],
\end{equation}
where $p(\y_k, \bar{\ppos} | \pvec, \sigma^2_k) = p(\y_k | \pvec, \sigma^2_k)p(\bar{\ppos} | \ppos)$. 

The data obtained up to epoch $k$ produce via \eqref{eq:ML} a sequence of estimates
$\check{\pvec}_1, \check{\pvec}_2, \dots, \check{\pvec}_k$. The
MSE-optimal combination of the estimates is formed using
weights based on the inverse covariance matrix for each estimate \cite{KailathEtAl2000_linear}. For epoch $k$, the latter is well approximated by the Fisher information matrix in \eqref{eq:FIM}, or by an estimate of it which we denote $\what{\FIM}_k$.  For notational simplicity let 
\begin{equation}
\check{\pvec}_0 = \begin{bmatrix} \0 \\ \bar{\ppos}
 \end{bmatrix} \quad \text{and} \quad
\what{\FIM}_0 =  \begin{bmatrix}
\0 & \0 \\
\0 & \IM_x
\end{bmatrix}
\end{equation}
denote the prior estimate and the corresponding information matrix, respectively.
Then we can compute the following linear combination recursively:
\begin{equation}\label{eq:linearcombiner}
\begin{split}
\what{\pvec}_k &= \left( \sum^k_{i=0} \what{\FIM}_i \right)^{-1} \left(\sum^k_{i=0} \what{\FIM}_i\check{\pvec}_i \right) =  \what{\IM}^{-1}_k \mbf{s}_k
\end{split}
\end{equation}
where
\begin{equation*}
\begin{cases}
\what{\IM}_k &= \what{\IM}_{k-1} + \what{\FIM}_k\\
\mbf{s}_{k} &= \mbf{s}_{k-1} + \what{\FIM}_k\check{\pvec}_k 
\end{cases} \quad \text{and} \quad
\begin{cases}
\what{\IM}_0 &= \what{\FIM}_0\\
\mbf{s}_{0} &= \what{\FIM}_0 \check{\pvec}_0.
\end{cases} 
\end{equation*}

\emph{Remark:} If a constant noise level $\sigma^2_0$ is used in \eqref{eq:FIM}, i.e.,
$\what{\FIM}_k = \FIM_k(\check{\ppos}_k,\sigma^2_0)$, then one can verify that
\eqref{eq:linearcombiner} is invariant to the nominal value $\sigma^2_0 >
0$. To make \eqref{eq:linearcombiner} robust with respect to noise
outliers the corresponding estimate of $\FIM_k \succ \0$ should decrease when there are outlying observations in epoch $k$. This
can be achieved using the estimated noise variance from
\eqref{eq:ML} for each epoch, which we denote
$\check{\sigma}^2_k$. More concretely, we use $\hat{\sigma}^2_k =
\max(\check{\sigma}^2_k,\sigma^2_0)$. In this way, the estimator adapts to
noise outliers that exceed a nominal $\sigma^2_0$ and at the same time
occasional overestimation of the information matrix is prevented when $\check{\sigma}^2_k$ is small.

\subsection{Minimization method}

We propose a computationally efficient gradient-based method to
solve \eqref{eq:ML}. First we note that the negative log-likelihood
can be expressed as
\begin{equation}\label{eq:NLL}
\begin{split}
-\ln p(\y_k, \bar{\ppos} | \pvec, \sigma^2_k) &=
\frac{\sigma^{-2}_k}{2} \| \y_k - \mbs{\mu}_k - \regclock_k \pclock - \frac{1}{c}\regpos_k\ranges(\ppos) \|^2_{\covnoise^{-1}_k}  \\
&\quad + \frac{n_k}{2} \ln \sigma^2_k + \frac{1}{2}\| \ppos - \bar{\ppos} \|^2_{\IM_x} + K,
\end{split}
\end{equation}
where $K$ is a constant. We will subsequently drop the subindex $k$
for notational convenience. The minimizing $\pclock$ and $\sigma^2$
can be expressed as 
\begin{equation}\label{eq:ML_clockplusnoise}
\begin{split}
\check{\pclock}(\ppos) &= \left(\regclock^\T \covnoise^{-1}  \regclock\right)^\dagger
\regclock^\T \covnoise^{-1} \left( \y-\mbs{\mu} -
c^{-1}\regpos\ranges(\ppos) \right) \\
\check{\sigma}^2(\ppos) &= \frac{1}{n}\left\| \proj \left(  \y - \mbs{\mu} - c^{-1}
\regpos \ranges(\ppos)  \right) \right\|^2_{\covnoise^{-1}} \\
\end{split}
\end{equation}
where
\begin{equation*}
\begin{split}
\proj \triangleq \I - \regclock ( \regclock^\T \covnoise^{-1}
\regclock)^\dagger \regclock^\T \covnoise^{-1}
\end{split}
\end{equation*}
is a projector matrix. After inserting \eqref{eq:ML_clockplusnoise}
into \eqref{eq:NLL}, the maximum likelihood estimate of $\ppos$ can be
obtained by solving
\begin{equation}\label{eq:V}
 \check{\ppos} = \argmin_{\ppos} \;  \underbrace{ \ln V_0(\ppos) + V_1(\ppos) }_{\triangleq V(\ppos)},
\end{equation}
where
\begin{equation}
\begin{split}
V_0(\ppos) =  \check{\sigma}^2(\ppos) \quad \text{and} \quad
V_1(\ppos) = \frac{1}{n} \| \ppos - \bar{\ppos} \|^2_{\IM_x} .
\end{split}
\end{equation}

The gradient of $V(\ppos)$ can be written as
\begin{equation*}
\begin{split}
 \partial V(\ppos) &= \frac{1}{V_0(\ppos)} \partial V_0(\ppos)
 +  \partial V_1(\ppos),
\end{split} 
\end{equation*}
where compact expressions of the gradients $\partial V_0(\ppos)$ and $\partial V_1(\ppos)$
are given in Appendix~\ref{app:A}. Starting from an initial point $\check{\ppos}^{0}$, we formulate a gradient descent method
\begin{equation}\label{eq:gradientmethod}
\begin{split}
 \check{\ppos}^{i+1} &= \check{\ppos}^{i} + \alpha_i \pdir_i,
\end{split}
\end{equation}
where
\begin{equation}\label{eq:direction}
\begin{split}
\pdir_i \triangleq -\frac{\partial V_0(\ppos) + V_0(\ppos) \partial
  V_1(\ppos) }{\left\| \partial V_0(\ppos) + V_0(\ppos) \partial
    V_1(\ppos) \right\|} \propto -  \partial V(\ppos)
\end{split}
\end{equation}
and the step size $\alpha_i$ is chosen by a line search
\begin{equation}\label{eq:linesearch}
\min_{\alpha_i \in I} \; V(\check{\ppos}^{i} + \alpha_i \pdir_i)
\end{equation}
in the interval $I = \left[0, \eta\|  \check{\ppos}^{i} -
  \check{\ppos}^{i-1} \|\right]$ where $\eta$ is a user
parameter which determines the upper limit on the step
size. When prior information is available the initial
point can be taken as $\check{\ppos}^0 = \bar{\ppos}$. If it is unavailable the centroid of the known transmitting node coordinates,
i.e. $\check{\ppos}^0 = \frac{1}{4} \sum_i \ppos_i$, or the estimate from a previous epoch can be used.

In summary, for each epoch, \eqref{eq:ML} is solved by iterating
\eqref{eq:gradientmethod} until convergence, followed by
insertion of the position estimate into \eqref{eq:ML_clockplusnoise}. Then a
optimal estimate of $\pvec$ is formed via \eqref{eq:linearcombiner}. A summarizing
pseudo-code is given in Algorithm~\ref{alg:iterative}.

\begin{algorithm}
  \caption{Online estimator at a generic epoch} \label{alg:iterative}
\begin{algorithmic}[1]
    \State Input: $\y$, $\mbf{s}$ and $\what{\IM}$
    \State Initialize $i=0$ and $\check{\ppos}^i$
    \Repeat
        \State Compute $\pdir_i$ via \eqref{eq:direction}
        \State Set $\alpha_i$ using \eqref{eq:linesearch}
        \State $\check{\ppos}^{i+1} = \check{\ppos}^{i} + \alpha_i
        \pdir_i$
        \State $i := i+1$
    \Until{ $\alpha_i < \epsilon $ }
    \State Compute $\check{\pclock}$ and $\check{\sigma}$ via
    \eqref{eq:ML_clockplusnoise}
    \State Compute $\what{\FIM}$ via \eqref{eq:FIM}
    \State $\IM := \IM + \what{\FIM}$
    \State $\mbf{s} := \mbf{s} + \what{\FIM}\check{\pvec}$
    \State $\what{\pvec} = \what{\IM}^{-1} \mbf{s}$
    \State Output: $\what{\pvec}$, $\mbf{s}$ and $\what{\IM}$
\end{algorithmic}
\end{algorithm}

\section{Numerical experiments}

We perform a numerical evaluation of \textsc{Swins}, comparing
the accuracy of the online estimator with the Cramér-Rao bounds. The root mean-square error (\textsc{Rmse}) of the parameter estimates was
computed using $10^3$ Monte Carlo simulations. 

In the following examples we set the unknown clock parameters to $T_m = 50 \times 10^{-9}$
and $T_u = 50 \times 10^{-9}$ [s]. The unknown $\phi_u$ contains
the time of flight and the offset $\Delta_1$ that we set to $5 \times 10^{-9}$ [s].
Note however that the bounds are invariant to these parameter values. 
The numbers of clock cycles were set to $M=100$ and to $N=101$. In all
examples the master is located at the following coordinates
\begin{equation*}
\ppos_m = \begin{bmatrix}
1 \\1 
\end{bmatrix}.
\end{equation*}

In the first scenario  we consider a situation in which we have prior
information about the position, modeled by the distribution
$\mathcal{N}(\bar{\ppos}, \IM^{-1}_x)$, and no additional transceivers
are present. In the second scenario, we consider no prior information
(i.e. $\IM_x = \0$) but add transceivers located at 
\begin{equation*}
\ppos_1 = \begin{bmatrix} 11 \\ 11 \end{bmatrix},
\ppos_2 = \begin{bmatrix} 1 \\ 11 \end{bmatrix} \text{ and }
\ppos_3 = \begin{bmatrix} 11 \\ 1 \end{bmatrix},
\end{equation*}
cf. the configuration in Fig.~\ref{fig:crb_phi}.

For the online estimator we set the nominal $\sigma_0$ to $10$~[ns]
and let the estimator adapt to noise outliers that exceed
$\sigma^2_0$. The upper limit on the relative step size, $\eta$, is
set to 1.2.  We set the tolerance
$\epsilon$ to $10^{-7}$.

\subsection{Master node and no transceivers}

In the first scenario, the prior information is given by
\begin{equation*}
\bar{\ppos} = 
\begin{bmatrix}
9 \\8
\end{bmatrix}
\quad \text{and} \quad
\IM_x = \sigma^{-2}_{x} \I_2,
\end{equation*}
where $\sigma_x$ parameterizes the precision of $\bar{\ppos}$ in
meters. The unknown position of the node is randomized as $\ppos \sim \mathcal{N}(\bar{\ppos}, \IM^{-1}_x)$.

\begin{figure*}
   \begin{center}
   \subfigure{
          \includegraphics[width=0.95\columnwidth]{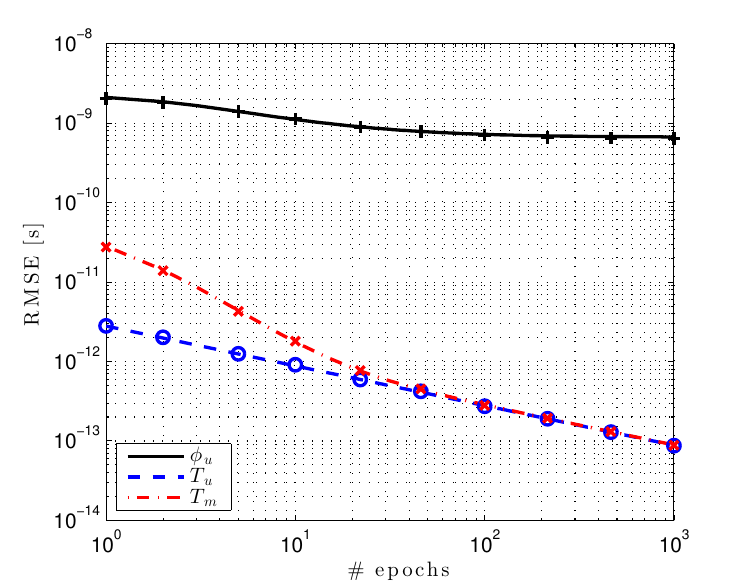}
   }
   \quad
   \subfigure{
          \includegraphics[width=0.95\columnwidth]{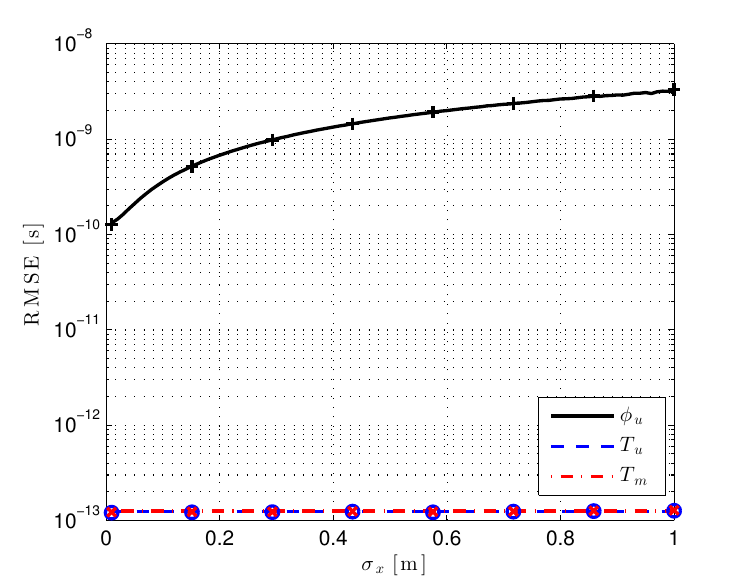}
   }
   \end{center}
   \caption{Master without transceivers: System resolution limits (\textsc{Hcrb} as lines) and estimator performance (crosses and circles) as a function of epochs (left) and precision of prior (right), respectively. The noise variance is $\sigma_k$ is fixed
    to 2~[ns]. (Left) $\sigma_x$ is 0.20~[m]. (Right) Number of epochs is 500.}
   \label{fig:hcrb_rmse}
   \end{figure*}

The resolution limits of \textsc{Swins}, given by the \textsc{Hcrb}
\eqref{eq:hcrb}, are shown in Fig.~\ref{fig:hcrb_rmse}. When
$\sigma_x$ is 20~[cm] and the measurement noise level $\sigma_k$ is
2~[ns], we note that the \textsc{Hcrb} of $\phi_u$ reaches sub-nanosecond
levels as the number of epochs $k$ increases. The bound of $T_m$
eventually collapses to that of $T_u$, whose accuracy is fundamentally
limited by the errors of the timing device,
cf.~\eqref{eq:measurement_Tu}. In this scenario the online estimator
achieves the \textsc{Hcrb} for all parameters. 

Fig.~\ref{fig:hcrb_rmse} illustrates also how the accuracy of the
initial position estimate $\bar{\ppos}$, namely $\sigma_x$, limits the accuracy of
$\phi_u$. For 500 epochs, a position accuracy about $\pm
50$~cm ($\sigma_x=0.25$) results in sub-nanosecond resolution limit
for $\phi_u$. The bounds for $T_m$ and $T_u$ are left virtually
unaffected by $\sigma_x$.

\subsection{Master node with three transceivers}

\begin{figure*}
   \begin{center}
   \subfigure{
          \includegraphics[width=0.95\columnwidth]{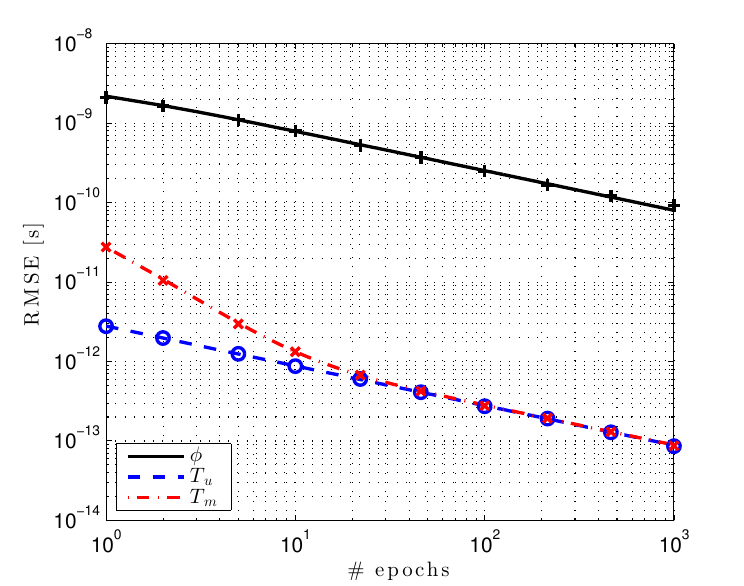}
   }
   \quad
   \subfigure{
          \includegraphics[width=0.95\columnwidth]{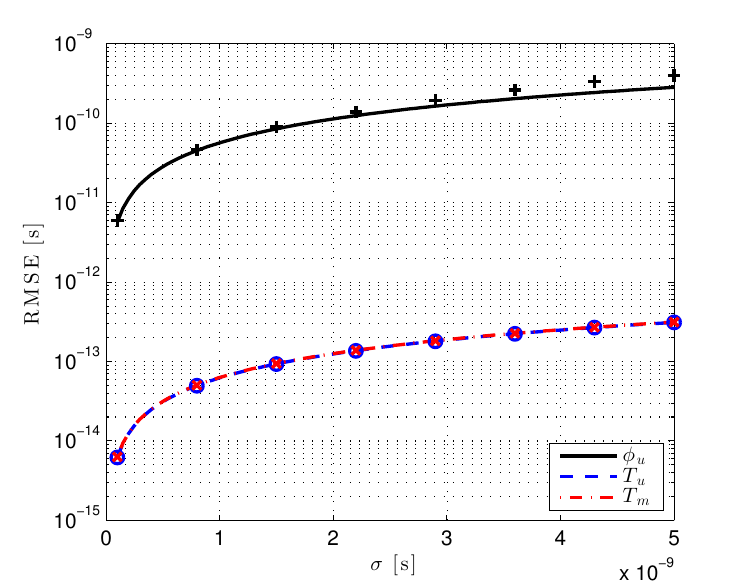}
   }
   \end{center}
   \caption{Master with transceivers: System resolution limits (\textsc{Crb} as lines) and estimator performance (crosses and circles) as a function of epochs (left) and noise level (right), respectively. (Left) $\sigma$ is 2~[ns]. (Right) Number of epochs is 500.}
   \label{fig:crb_rmse}
   \end{figure*}

The unknown position of the node is fixed at $\ppos = [9 \;
8]^\top$. The resolution limits of \textsc{Swins}, given by the
\textsc{Crb} in \eqref{eq:crb}, are shown in Fig.~\ref{fig:crb_rmse}. For a
noise level of $\sigma_k =$ 2~[ns], the \textsc{crb} of $\phi_u$ reaches sub-nanosecond levels already at 10 epochs.
Similar to the previous scenario the online estimator
attains the bounds, which now decrease steadily with the number of epochs.

Fig.~\ref{fig:crb_rmse} illustrates also how the measurement noise
level limits the accuracy of $\phi_u$.  The estimation errors decrease
as the unknown noise decreases $\sigma_k \rightarrow 0$. A small gap
to the \textsc{crb} for $\phi_u$ is visible when the noise level increases to 5 [ns].

\section{Conclusion}

We have designed a  scalable system, denoted \textsc{Swins}, in which an indefinite number of receiving wireless units can synchronize to a single master
clock. The synchronization is performed at the level of discrete clock ticks and the mechanism can be implemented with passive receivers, thereby obviating the need for two-way communication and time-stamp exchanges.

By deriving Cramer-Rao
bounds for the data model we can conclude that \textsc{Swins} advances the limits wireless synchronization towards sub-nanoseconds levels based on state-of-the art hardware components. An online estimator based on the maximum likelihood approach was also developed that can operate
with prior position information or, when such information is absent,
with the proposed positioning infrastructure. The numerical
experiments show that the estimator is statistically efficient.

In future work we will consider applications which can benefit from precise timing information and, furthermore, study the impact on performance of the geometric configuration of the transmitting nodes. 


\appendices
\section{Derivation of gradient}\label{app:A}

The gradient of $V_1$ is readily obtained as
\begin{equation}\label{eq:dV1}
\begin{split}
\partial V_1 &= \frac{2}{n}\IM_x (\ppos - \bar{\ppos}) .
\end{split}
\end{equation}
Due to the logarithm $\ln V_0$, we can equivalently redefine $V_0$ as $V_0 = n \check{\sigma}^2$. Then to obtain the gradient of $V_0(\ppos)$ we first re-write the function as
\begin{equation}\label{eq:V0_alt1}
\begin{split}
V_0 = \ranges^\T \W \ranges - 2\vvec^\top \ranges  +  (\y - \mbs{\mu})^\T\covnoise^{-1} \proj  (\y - \mbs{\mu}),
\end{split}
\end{equation}
where 
\begin{equation*}
\begin{split}
\W &=c^{-2} \regpos^\T \covnoise^{-1} \proj \regpos \\
\vvec &= c^{-1}\regpos^\top \covnoise^{-1} \proj  (\y - \mbs{\mu}).
\end{split}
\end{equation*}
Because \eqref{eq:V0_alt1} equals
\begin{equation*}
\begin{split}
V_0 &= \sum_i \sum_j [\W]_{ij} \rho_i \rho_j - 2 \sum_i w_i \rho_i + K,
\end{split}
\end{equation*}
where $K$ is a constant, the gradient can be expressed as 
\begin{equation}\label{eq:dV0}
\partial V_0= \sum_i \sum_j [\W]_{ij} \left( \grange_i\rho_j  + \rho_i
  \grange_j \right)- 2 \sum_i w_i \grange_i,
\end{equation}
where
\begin{equation*}
\begin{split}
\grange_i &\triangleq \pder_x \rho_i = \pder_x( \| \ppos - \ppos_i \|^2 )^{1/2} = \frac{\ppos - \ppos_i }{\| \ppos -  \ppos_i \|}.
\end{split}
\end{equation*}
The gradients in \eqref{eq:dV0} and \eqref{eq:dV1} are used in \eqref{eq:direction}.

\bibliographystyle{IEEEtran}
\bibliography{ref_sync}

\end{document}